\documentclass[preprint,prd,nofootinbib,tightenlines,amsmath,showpacs]{revtex4}

\usepackage{epsfig}
\usepackage{graphics,color}      
\usepackage{verbatim}
\usepackage{amsmath}    
\catcode`\@=11
\def\sla@#1#2#3#4#5{{%
 \setbox\z@\hbox{$\m@th#4#5$}%
 \setbox\tw@\hbox{$\m@th#4#1$}%
 \dimen4\wd\ifdim\wd\z@<\wd\tw@\tw@\else\z@\fi
 \dimen@\ht\tw@
 \advance\dimen@-\dp\tw@ \advance\dimen@-\ht\z@
 \advance\dimen@\dp\z@
 \divide\dimen@\tw@ \advance\dimen@-#3\ht\tw@
 \advance\dimen@-#3\dp\tw@ \dimen@ii#2\wd\z@
 \raise-\dimen@\hbox to\dimen4{%
 \hss\kern\dimen@ii\box\tw@\kern-\dimen@ii\hss}%
 \llap{\hbox to\dimen4{\hss\box\z@\hss}}}}

\def\slashed#1{%
 \expandafter\ifx\csname sla@\string#1\endcsname\relax
{\mathpalette{\sla@/00}{#1}}
\fi}
\def\declareslashed#1#2#3#4#5{%
 \expandafter\def\csname sla@\string#5\endcsname{%
#1{\mathpalette{\sla@{#2}{#3}{#4}}{#5}}}}
 \catcode`\@=12
\declareslashed{}{/}{.08}{0}{D}
 \declareslashed{}{/}{.1}{0}{A}
 \declareslashed{}{/}{0}{-.05}{k}
 \declareslashed{}{/}{.1}{0}{\partial}
 \declareslashed{}{\not}{-.6}{0}{f}

\def\hq#1#2{H^{(#1)}_{#2}}
\def\hqb#1#2{H^{(\bar{#1})}_{#2}}
\def\hbq#1#2{\bar{H}^{(#1)}_{#2}}
\def\hbqb#1#2{\bar{H}^{(\bar{#1})}_{#2}}
\def\lsim{\mathrel {\vcenter {\baselineskip 0pt \kern 0pt
    \hbox{$<$} \kern 0pt \hbox{$\sim$} }}}
\def\gsim{\mathrel {\vcenter {\baselineskip 0pt \kern 0pt
    \hbox{$>$} \kern 0pt \hbox{$\sim$} }}}

\begin{document}

\baselineskip=15pt

\preprint{hep-ph/0606065}

\hspace*{\fill} $\hphantom{-}$

\title{Weak Radiative $B$ Decays}

\author{Oleg Antipin and G. Valencia}

\email{oaanti02@iastate.edu}
\email[]{valencia@iastate.edu}

\affiliation{Department of Physics and Astronomy, Iowa State University, Ames, IA 50011\\}

\date{\today}

\begin{abstract}

We study the (tree-level) weak radiative decays of $B$ mesons. 
We present a numerical estimate for the inclusive $b \to X_c \gamma (\gamma)$ modes based on the free-quark decay. We then review what is known for the $B \to D^\star \gamma$ modes in the framework  of heavy quark effective theory and chiral perturbation theory. Finally, we extend these ideas to the double radiative decay
modes $B \to D \gamma \gamma$. We find that the $b \to X_c \gamma \gamma$ rate is about an order of magnitude larger than the corresponding $b \to X_s \gamma \gamma$ rate. We also find the branching ratio for the $B \to D \gamma \gamma$ mode with most favorable CKM angles at the few $\times10^{-8}$ level, comparable to predictions for $B \to K \gamma \gamma$.

\end{abstract}


\maketitle    
\section{Introduction}

The radiative penguin $b$  decay of the form $b \to s \gamma$ 
 has received much attention in the literature because it is sensitive to certain types of physics beyond the standard model. The HFAG quotes an average for the measured branching ratio $B(b\to s\gamma) = (354^{+30}_{-28}) \times 10^{-6}$ \cite{Barberio:2006bi}. Recently the double radiative decay mode $b\to s\gamma\gamma$ has received some attention in connection with the possibility of measuring it at a Super-B factory at the $10^{-7}$ level \cite{Hewett:2004tv}. This is the level at which it is expected to occur in the standard model 
 \cite{Soni:1997my,Reina:1997my}.

There is another type of radiative $b$ decay modes in which the weak decay proceeds via the charged current at tree-level, $b \to X_c \gamma (\gamma)$ and $b \to X_u \gamma (\gamma)$. These modes are expected to be dominated by standard model physics and have received much less attention. We begin this paper with a simple numerical estimate for these modes suggesting that $b \to X_c \gamma \gamma$ can also be observed at a Super-B factory. We then turn our attention to some of the exclusive modes.

The charged mode with one photon, $B^+ \to D^{+\star} \gamma$, has been studied before in the context of heavy quark effective theory as a potential probe for $V_{ub}$ by Grinstein and Lebed \cite{Grinstein:1999qr}. The neutral mode with one photon, $\bar{B}^0 \to D^{0\star} \gamma$, has been studied by Cheng {\it et. al.} \cite{Cheng:1994kp} in a slightly different context. We first review what is known about these modes using the framework of heavy quark effective theory and chiral perturbation theory and present new results that include the effect of intermediate positive parity states. We find that these intermediate states have an important effect on the overall rates, particularly for the charged modes, due to a partial cancellation that occurs in the leading order amplitude.

We then extend these results to the case of the double radiative decay modes $B \to D \gamma\gamma$. We include the lowest lying positive parity states in our calculation and find that they play an important role, just as they do in the single radiative decay modes.  The double radiative decay modes present, in principle, the opportunity to study the heavy quark and chiral expansions by looking at different kinematic regions. In practice, however, this may not be possible due to the small rates. 
We find that the mode with the most favorable CKM angles can occur at the few $\times 10^{-8}$ level and is perhaps observable at a Super-B factory.

\section{$b\to X_c \gamma$ and $b\to X_c \gamma \gamma$}
 
We begin by discussing the inclusive radiative decays. We work in the free quark approximation where $b\to X_c \gamma$ and $b\to X_c \gamma \gamma$ arise from the tree level quark processes $b \to c \bar{u} d \gamma$ 
and $b \to c \bar{u} d \gamma\gamma$ (we take $V_{ud}=1$).  
Our goal is  to compare these processes to the one-loop processes $b \to X_s \gamma$ and $b \to X_s \gamma \gamma$ that have been studied at length in the literature. We will limit ourselves to a numerical estimate with the aid of CompHEP \cite{comphepre}.

\begin{figure}[htb]
\vspace{0.5in}
\includegraphics[width=4 in]{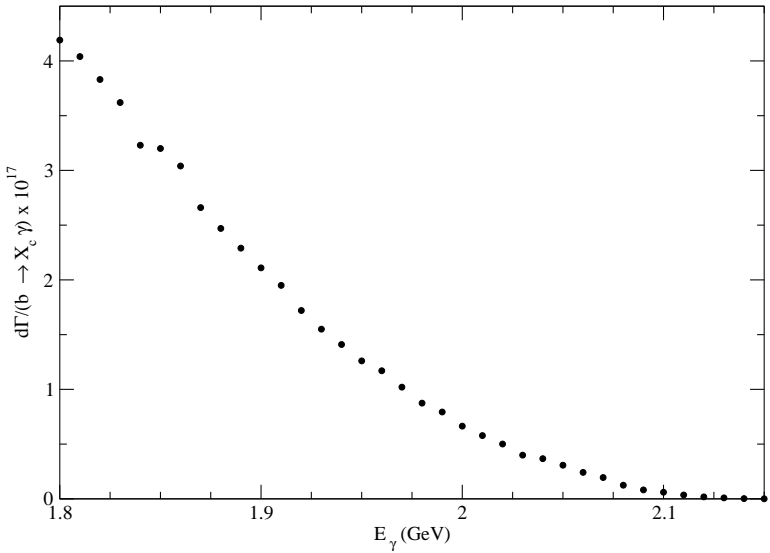}
\centering
\caption{Photon energy spectrum in $b\to X_c \gamma$ with $\theta_{min}=5^\circ$.}
\label{f:bsg}
\end{figure}

The process $b \to X_c \gamma$ is to be compared with the inclusive $b \to X_s \gamma$. Accordingly we impose a cut on the photon energy $1.8$~GeV~$ \leq E_\gamma \leq$~$2.8$~GeV corresponding to the energy range studied by the Belle collaboration 
\cite{Koppenburg:2004fz}. We also impose a separation cut that requires a minimum angle $\theta_{min}$ between the photon  and the final state quarks. With values of $5^\circ\leq \theta_{min}\leq 20^\circ$ for this minimum angle we obtain rates $\Gamma(b\to c\bar{u} d \gamma)$ between $3\times 10^{-18}$~GeV and $5\times 10^{-18}$~GeV. To estimate the branching ratio for the inclusive process we then use
\begin{eqnarray}
B(b\to X_c \gamma) &=& \frac{\Gamma(b\to c\bar{u} d \gamma)_{th}}{\Gamma(b\to c e^- \nu)_{th}}B(B^+\to X_c e^+ \nu)_{exp}\nonumber \\
&\sim & (7 - 11)\times 10^{-6}.
\label{incg}
\end{eqnarray}
To obtain this number we used $V_{cb}=0.0413$ \cite{Eidelman:2004wy}, $B(B^+\to X_c e^+ \nu)_{exp} = (11.15\pm0.26\pm0.41)\%$ \cite{Okabe:2004fm} and quark mass values $m_b=4.8$~GeV, $m_c=1.5$~GeV which are the ones used in the theory estimates of $b\to X_s \gamma\gamma$. We show in Figure~\ref{f:bsg} the photon energy spectrum for the case of $\theta_{min}=5^\circ$. The characteristic bremsstrahlung spectrum falls rapidly with the photon energy and for this reason $B(b\to X_c \gamma)$ is much smaller than the penguin mode $B(b \to X_s \gamma) = (3.3\pm0.4)\times10^{-4}$ \cite{Eidelman:2004wy} in this energy range.

Next we consider the double radiative mode $B\to X_c \gamma \gamma$ in a similar way. We wish to compare it to the penguin process $B\to X_s \gamma\gamma$ which has not been observed. We use instead the theoretical calculation of Reina {\it et. al.} 
\cite{Soni:1997my,Reina:1997my} which finds $B(B\to X_s \gamma\gamma)\sim 3.7 \times 10^{-7}$ (including LO QCD corrections and using the quark masses mentioned above). They also require that the energy of each photon be larger than $0.1~{\rm GeV}$, that the photon pair invariant mass be larger than $0.1m_b$, and that the photons be separated from each other and from final state quarks by at least $20^\circ$. They find a  spectrum that is sharply peaked at low $M_{\gamma\gamma}$ invariant mass. 

\begin{figure}[htb]
\vspace{0.5in}
\includegraphics[width=4 in]{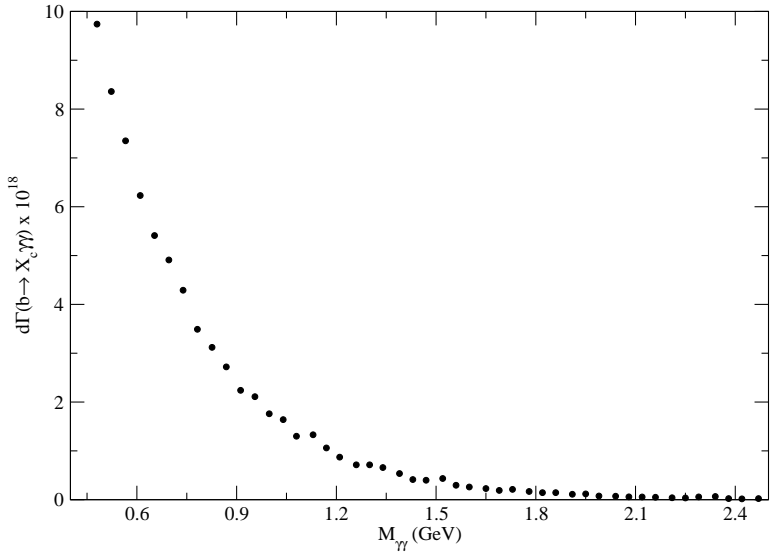}
\centering
\caption{$M_{\gamma \gamma}^2 $ distribution for the process $b\to X_c \gamma \gamma$ with the cuts described in the text.}
\label{f:bcgg}
\end{figure}

For our calculation we estimate  $B\to X_c \gamma \gamma$ from the leading tree-level contribution $b \to c \bar{u} d \gamma\gamma$.
With the same cuts used by Ref.~\cite{Soni:1997my,Reina:1997my} for $B\to X_s \gamma\gamma$ we obtain $\Gamma(b \to c \bar{u} d \gamma\gamma)=1.9 \times 10^{-18}$~GeV and this goes up to $3.2\times 10^{-18}$~GeV if the angular cuts are relaxed to $10^\circ$. 
Proceeding as in Eq.~(\ref{incg}) we arrive at
\begin{equation}
B(B\to X_c \gamma \gamma) \sim (4.2-7.2) \times 10^{-6}.
\end{equation} 
In Figure~\ref{f:bcgg} we show the two photon invariant mass distribution. 
 
Once again we see a spectrum that is strongly peaked at low invariant mass. Comparing our results to those of Ref.~\cite{Soni:1997my,Reina:1997my} it is clear that $B\to X_c \gamma \gamma$ can be a significant background to $B\to X_s \gamma \gamma$. 
Notice that the total rate for the double radiative decay mode is comparable to that for $b\to X_c \gamma$. This of course is due to the much tighter photon energy cuts imposed in that case.

\section{Exclusive Modes and HQET}

In this section we collect the ingredients necessary to calculate  the  amplitudes for the exclusive modes guided by heavy quark effective theory, combined with SU(3) chiral perturbation theory \footnote {Similar methods were used in Ref.\cite{Fajfer:2001ad} to calculate rare $D^0$ decay to two photons}. 

\subsection{Radiative decays of heavy mesons}

The strong interactions involving the heavy meson $(0^-,1^-)$ doublet 
(the lightest pseudoscalar and vector mesons containing one heavy quark) and light pseudo-scalars are described by the effective Lagrangian (we drop the subscript $v$ that indicates the velocity of the heavy meson to simplify the notation) \cite{Wise:1992hn,Burdman:1992gh,manoharbook}
\begin{equation}
{\cal L}=-i Tr(\hbq{Q}{a} v \cdot D_{ba}\hq{Q}{b})+g Tr(\hbq{Q}{a} \hq{Q}{b} \gamma_{\nu} \gamma_{5} A_{ba}^{\nu}).
\label{hqetst}
\end{equation}
In Eq.~(\ref{hqetst}) we use the standard notation in which 
\begin{itemize}
\item The heavy pseudoscalar and vector meson fields with heavy quark $Q$ and light anti-quark $\bar{q}_a$ are destroyed and created  by the field $\hq{Q}{a}$  and its Hermitian conjuagte. They are given by 
\begin{equation}
\hq{Q}{a}=\frac{1+\slashed{v}}{2}\left(P^{(Q)\star}_{a\mu}\gamma^\mu-P^{(Q)}_a\gamma_5\right)
\label{hfield}
\end{equation}
which transforms under chiral symmetry as $\hq{Q}{a}\to\hq{Q}{b}U^\dagger_{ba}$. The Hermitian conjugate matrix $\hbq{Q}{a}=\gamma_0 H_a^{(Q)\dagger} \gamma_0$.

\item Mesons containing heavy anti-quarks $\bar{Q}$ and light quarks $q_a$ are destroyed and created  by the fields
\begin{eqnarray}
\hqb{Q}{a} &=& (P^{(\bar{Q})\star}_{a\mu}\gamma^\mu-P^{(\bar{Q})}_a\gamma_5) \frac{1-\slashed{v}}{2} \nonumber \\
\hbqb{Q}{a} &=& \gamma_0 H_a^{(\bar{Q})\dagger} \gamma_0 
\end{eqnarray}

\item The pseudo-Goldstone boson octet, $\phi$ is incorporated into a $3\times 3$ unitary matrix $\Sigma=\exp\left(2i\phi/f_\pi\right)$ which transforms under chiral symmetry as $\Sigma \to L \Sigma R^\dagger$. In Eq.~(\ref{hqetst}) they enter through the matrix $\xi$ where $\Sigma =\xi^2$ and the transformation properties of $\xi$ under chiral symmetry $\xi \to L \xi U^\dagger = U \xi R^\dagger$ define the matrix $U$. The charges of the light quarks appear through the diagonal matrix ${\cal Q}$ with entries $2/3,-1/3,-1/3$.

\item The chiral covariant derivative and axial current in Eq.~(\ref{hqetst}) are given by 
\begin{eqnarray}
D_{ab}^{\mu} &=& \delta_{ab}\partial^{\mu}- V_{ab}^{\mu}=\delta_{ab}\partial^{\mu}-\frac{1}{2}(\xi^{\dagger}\partial^{\mu}\xi+\xi \partial^{\mu}\xi^{\dagger})_{ab}     \nonumber \\
A^\mu_{ab} &=& \frac{i}{2}(\xi^{\dagger}\partial^{\mu}\xi-\xi \partial^{\mu}\xi^{\dagger})_{ab}
\end{eqnarray}

\end{itemize}

The leading order electromagnetic coupling is obtained from Eq.~(\ref{hqetst}) by minimal substitution. However, the couplings arising from this procedure (for charged $B$ and $D$ mesons) do not contribute to the processes $B\to D \gamma$ or  $B \rightarrow D \gamma \gamma $ as can be seen by explicit computation. The lowest order electromagnetic coupling that will contribute to these processes is the transition magnetic moment \cite{Cho:1992nt,Amundson:1992yp,Casalbuoni:1996pg,manoharbook,Stewart:1998ke}. For mesons containing a heavy quark it can be written as,
\begin{equation}
{\cal L}_{em}=-\frac{ee_Q\mu^{(h)}}{4} Tr(\hbq{Q}{a}\sigma_{\mu\nu} \hq{Q}{a} )  F^{\mu\nu}-\frac{e \mu^{(l)}}{4} Tr(\hbq{Q}{a} \hq{Q}{b} \sigma_{\mu\nu}{\cal Q}^{\xi ba}  F^{\mu\nu}) ,
\label{emmag}
\end{equation}

where ${\cal Q}^{\xi}=\frac{1}{2}(\xi^{\dagger}{\cal Q}\xi+\xi{\cal Q}\xi^{\dagger})$.

The coupling consists of two terms corresponding to the decomposition of the electromagnetic current into heavy and light quark parts, 
\begin{equation}
\mu_a=e_Q\mu^{(h)}+e_a\mu^{(l)}\equiv \mu_a^{(h)}+\mu_a^{(l)}=\frac{e_Q}{\Lambda_Q}+\frac{e_a}{\Lambda_a},
\end{equation}
where $e_a$ is the charge of the light-quark. 
The heavy quark contribution at leading order in the $1/m_Q$ expansion is  given by $\Lambda_Q=m_Q$ \cite{Isgur:1989ed,Isgur:1989vq}. The light quark contribution in the SU(3)  symmetry limit is usually called $\Lambda_a^{-1}=\beta$ \cite{manoharbook}. This constant has been estimated in vector meson dominance models \cite{Colangelo:1993zq} as well as in chiral quark models \cite{Colangelo:1994jc} (along with SU(3) breaking corrections). 
The leading SU(3) violations have also been calculated \cite{Amundson:1992yp}. More generally, the heavy quark contribution when the mesons have different velocity is modified to $\mu_a^{(h)}=e_Q\xi(\omega)/m_Q$ \cite{Casalbuoni:1996pg,Isgur:1990jf} where $\omega=v\cdot v^\prime$ and $\xi(\omega)$ is the Isgur and Wise function.

Eq.~(\ref{emmag}) generates the following amplitudes,
\begin{eqnarray}
{\cal M}(D^\star(\eta) \rightarrow D \gamma(q,\epsilon)) &=&
-ie \mu_D \epsilon^{\mu\nu\alpha\beta}\epsilon^\star_{\mu} \eta_{\nu}q_{\alpha}v_{\beta}, \nonumber \\
{\cal M}(D^\star(\eta_1) \rightarrow D^\star(\eta_2) \gamma(q,\epsilon)) &=& e\mu_{D\star} \left( q\cdot\eta_1\epsilon^\star\cdot\eta^\star_2- q\cdot\eta^\star_2\epsilon^\star\cdot\eta_1\right),
\label{hrad}
\end{eqnarray}
where we have defined $\mu_{D\star}\equiv ( \mu_a^{D(h)}-\mu_a^{D(l)})$. Analogous relations with $\mu_{D,D\star}\to\mu_{B,B\star}$ then hold for the  $B$ system. 
In the heavy quark and SU(3) limits, the magnetic moments consist only of the light quark contribution given by $\mu_a=e_a\beta$. For our numerical estimates we will use the leading magnetic moments as well as the magnetic  moments in three models tabulated in Ref.~\cite{Casalbuoni:1996pg}: ``$\chi$LM'' a chiral loop model; ``VMD'' a vector meson dominance model; and ``RQM'' a relativistic quark model.

The magnetic moments from Eq.~(\ref{emmag}) are defined for on-shell transitions between a vector and a pseudo-scalar of the same mass. In our calculations one of the mesons will be off its mass shell and the corresponding form-factors  should be evaluated at $k^2\sim -\delta m^2$ for the single radiative decay modes: $e\mu\equiv g_M(0) \to e\mu g_M(-\delta m^2)$ where $\delta m \equiv m_b-m_c$. 
On general grounds one expects the form factor to change over a characteristic scale $\Lambda_{QCD}$, so there is large uncertainty associated with the use of the on-shell form factors. Formally the results we obtain, correspond to the  so called generally low velocity (GL) limit in which \cite{Grinstein:1999qr}
\begin{equation}
\delta m \sim \Lambda_{QCD} \ll m_b.
\end{equation}
Alternatively one can model the momentum dependence of the form factors as was done in Ref.~\cite{Cheng:1994kp} for the neutral modes. We will not include a momentum dependence of the form factors in our estimates but instead introduce two additional effects.  
First we will keep certain terms that are formally of order $\delta m/m_b$ arising from spin one propagators as described later on.  We will also consider additional heavy meson intermediate states; the positive parity P-waves of the system Q$\bar{q}$. However, we will neglect higher total spin resonances. 

The positive parity states that we include are predicted  by HQET to lie in  two distinct multiplets: 
($0^+,1^+$)=($B_0,B_1$) and ($1^+,2^+$)=($\tilde{B_1},B_2$). Generically we will refer to them as $B^{\star\star}$ and will discuss the case of $B$ mesons for definiteness, with corresponding results for $D$ mesons also used in our calculation. 
The velocity dependent fields are introduced in a manner similar to the field $H$ in Eq.~(\ref{hfield}) \cite{Falk:1991nq,Kilian:1992hq}
\begin{eqnarray}
S&=&\frac{1}{2}(1+\slashed{v})[\slashed{B_1}\gamma_{5}-B_0 ]
\nonumber \\
T^{\mu}&=&\frac{1}{2}(1+\slashed{v})\left[B_2^{\mu\nu}\gamma_{\nu}-\sqrt{3/2}\tilde{B}_{1\nu}\gamma_{5}(g^{\mu\nu}-\frac{1}{3}
\gamma^{\nu}(\gamma^{\mu}-v^{\mu})) \right]
\label{stfields}
\end{eqnarray}
The electromagnetic transitions between a member of these doublets and a member of the  ($0^-, 1^-$) doublet have also been discussed in the literature. Analogously to Eq.~(\ref{emmag}) the couplings receive contributions from the heavy and light quark currents. 

The heavy quark contribution to the ($1^+,2^+$) to ($0^-, 1^-$) transition in the charm case can be found in \cite{Korner:1992pz}. A simple way to reproduce those results consists of using the matrix elements for $<B^{\star\star}|V^\mu|B>$ obtained by Isgur and Wise \cite{Isgur:1990jf} to match the effective Lagrangian ${\cal L}\sim Tr\left(\bar{H}_i\gamma^\beta T_j^\alpha {\cal Q}_{ij} F_{\alpha\beta}\right)$.  
Similarly one can start from the Isgur and Wise results and impose gauge invariance for the kinematic conditions we consider to obtain the heavy quark contribution to the ($0^+,1^+$) to ($0^-, 1^-$) transition.  

The light quark contributions can be determined at leading order in chiral perturbation theory in terms of two unknown constants,
\begin{eqnarray}
{\cal L}&=&-\frac{ie}{\Lambda^\prime_{3/2}}Tr\left(\bar{H}_iT_j^\alpha\gamma^\beta {\cal Q}_{ij} F_{\alpha\beta}\right)
-\frac{e}{4\Lambda^\prime_{1/2}} Tr\left(\bar{H}_iS_j\sigma_{\mu\nu}{\cal Q}_{ij}F^{\mu\nu}\right) + {\rm ~h.~c.}
\end{eqnarray}

Combining the heavy and light quark contributions for the $T$ doublet  leads to the amplitudes (we include only those that do not vanish at leading order in the $1/m_Q$ and chiral expansions)
\begin{eqnarray}
{\cal M}(B_2(v,\epsilon_2)\to B^\star(v,\epsilon_V)\gamma(q,\epsilon)) &=& -2\sqrt{3}e\mu^T_B\left(\epsilon_2^{\alpha\beta}q_\alpha\epsilon^\star_{V\beta}\epsilon^\star\cdot v -\epsilon_2^{\mu\beta}\epsilon^\star_{V\beta}\epsilon^\star_\mu q\cdot v \right) \nonumber \\
{\cal M}(\tilde{B}_1(v,\epsilon_1)\to B^\star(v,\epsilon_V)\gamma(q,\epsilon)) &=& i\sqrt{2}e\mu^T_B\epsilon^{\mu\nu\alpha\beta}
q_\mu\epsilon^\star_{\nu}\epsilon^\star_{V\alpha}\epsilon_{1\beta}  \nonumber \\
{\cal M}(\tilde{B}_1(v,\epsilon_1)\to B(v)\gamma(q,\epsilon)) &=& 
2\sqrt{2}e\mu^T_B\left(q\cdot\epsilon_1 v\cdot\epsilon^\star-q\cdot v\epsilon_1\cdot \epsilon^\star\right).
\label{bssemm}
\end{eqnarray}
Similarly for the $S$ doublet we obtain
\begin{eqnarray}
{\cal M}(B_1(v,\epsilon_1)\to B^\star(v,\epsilon_V)\gamma(q,\epsilon)) &=& -ie\mu^S_B
\epsilon^{\mu\nu\alpha\beta}
q_\mu\epsilon^\star_{\nu}\epsilon^\star_{V\alpha}\epsilon_{1\beta}  
 \nonumber \\
{\cal M}(B_1(v,\epsilon_1)\to B(v)\gamma(q,\epsilon)) &=& 
e\mu^S_B\left(q\cdot\epsilon_1 v\cdot\epsilon^\star-q\cdot v\epsilon_1\cdot \epsilon^\star\right)
 \nonumber \\
{\cal M}(B_0(v)\to B^\star(v,\epsilon_V)\gamma(q,\epsilon)) &=& e\mu^S_B
\left(q\cdot\epsilon^\star_V v\cdot\epsilon^\star-q\cdot v\epsilon^\star_V\cdot \epsilon^\star\right).
\label{bssemms}
\end{eqnarray}
Corresponding expressions for charm are obtained with the obvious replacements $\mu^{T,S}_B\to\mu^{T,S}_D$. For bottom
these effective coupling constants are
\begin{eqnarray}
\mu^T_B \equiv  \left(\frac{e_b\tau^{3/2}(1)}{m_b}+\frac{e_a}{\Lambda^\prime_{3/2}}\right), &&
\mu^S_B \equiv  \left(\frac{2e_b\tau^{1/2}(1)}{m_b}+\frac{e_a}{\Lambda^\prime_{1/2}}\right).
\end{eqnarray}
For the light quark contributions, $\Lambda^\prime_{3/2}$ corresponds to $\sqrt{3}/(f-f^\prime)$ of Ref.~\cite{Korner:1992pz}. Using their estimate for the $D$ system, we take $\Lambda^\prime_{3/2}\sim (2.75-3.5)$~GeV. Similarly, $\Lambda^\prime_{1/2}$ corresponds to $\Lambda^\prime$ of Ref.~\cite{Colangelo:1995sm} where it is estimated that $\Lambda^\prime_{1/2}\sim 1.25$~GeV. For the heavy quark contributions we have used the Isgur-Wise functions $\tau^{1/2,3/2}$ estimated in Ref.~\cite{Colangelo:1992kc} to be  $\tau^{1/2,3/2}(1)\sim 0.24$ ($\xi^{3/2}$ of Ref.~\cite{Korner:1992pz} corresponds to $\sqrt{3}\tau^{3/2}$) . More recent estimates of these functions using the light-front formalism \cite{Cheng:2003sm} and lattice calculations \cite{Becirevic:2004ta} indicate larger values, particularly for $\tau^{3/2}(1)$ up to $0.61$. These estimates also indicate that these functions vary by less than factors of two over the range of $\omega$ that is kinematically allowed in $B\to D \gamma\gamma$.

Finally we will also need ($0^+,1^+$) to ($0^+, 1^+$) and ($0^+,1^+$) to ($1^+, 2^+$) electromagnetic transitions. The former receives heavy and light quark contributions, whereas the latter only receives light quark contributions at order $1/m_Q$. They can be obtained from the Lagrangian
\begin{equation}
{\cal L}= -\frac{ee_Q}{4 m_Q} Tr(\bar{S}_{a}\sigma_{\mu\nu} S_{a} )  F^{\mu\nu}-\frac{e }{4\tilde{\Lambda}_{1/2}} Tr(\bar{S}_{a} S_{b} \sigma_{\mu\nu}{\cal Q}^{ab}  F^{\mu\nu}) -\frac{ie}{\tilde{\Lambda}_{3/2}}Tr\left(\bar{S}_iT_j^\alpha\gamma^\beta {\cal Q}_{ij} F_{\alpha\beta}\right).
\end{equation}
The new constants $\tilde{\Lambda}_{1/2,3/2}$ are not known. For our numerical estimates we use $\tilde{\Lambda}_{1/2}\sim \Lambda_a$ and $\tilde{\Lambda}_{3/2}\sim \Lambda_{3/2}$. With this choice the magnetic transitions within the $S$ multiplet are the same as those between members of the $H$ multiplet, for charm for example one has
\begin{eqnarray}
{\cal M}(D_1(\eta) \rightarrow D_0 \gamma(q,\epsilon)) &=&
-ie \mu_D \epsilon^{\mu\nu\alpha\beta}\epsilon^\star_{\mu} \eta_{\nu}q_{\alpha}v_{\beta}, \nonumber \\
{\cal M}(D_1(\eta_1) \rightarrow D_1(\eta_2) \gamma(q,\epsilon)) &=& e\mu_{D\star} \left( q\cdot\eta_1\epsilon^\star\cdot\eta^\star_2- q\cdot\eta^\star_2\epsilon^\star\cdot\eta_1\right),
\label{srad}
\end{eqnarray}
and corresponding expressions for bottom. The ($0^+,1^+$) to ($1^+, 2^+$) vertices for bottom are
\begin{eqnarray}
{\cal M}(B_2(v,\epsilon_2)\to B_1(v,\epsilon_V)\gamma(q,\epsilon)) &=& i2e\mu^{TS}_B\left(
\epsilon^{\alpha\mu\nu\gamma}\epsilon_2^{\alpha\beta}q_\beta v_\mu
\epsilon^\star_\nu \epsilon^\star_{1\gamma}+\epsilon^{\alpha\mu\nu\gamma}\epsilon_2^{\alpha\beta}q_\mu v_\nu
\epsilon^\star_\beta \epsilon^\star_{1\gamma}
\right)\nonumber \\
{\cal M}(\tilde{B}_1(v,\epsilon_i)\to B_1(v,\epsilon_f)\gamma(q,\epsilon)) &=&-e\mu^{TS}_B\sqrt{\frac{2}{3}}\left(
q\cdot \epsilon_i \epsilon^\star \cdot \epsilon^\star_f -
q\cdot \epsilon^\star_f\epsilon^\star \cdot \epsilon_i \right)
\nonumber \\
{\cal M}(\tilde{B}_1(v,\epsilon_1)\to B_0(v)\gamma(q,\epsilon)) &=& i2e\mu^{TS}_B\sqrt{\frac{2}{3}}
\epsilon^{\alpha\mu\nu\gamma}q_\alpha v_\mu \epsilon^\star_\nu \epsilon_{1\gamma},
\label{ttosem}
\end{eqnarray}
and corresponding expressions for charm. 
We have defined
\begin{equation}
\mu^{TS}_B\equiv \frac{e_a}{\tilde{\Lambda}_{3/2}}.
\end{equation}

\subsection{Weak transitions}

Within the standard model, the effective weak Hamiltonian responsible for the $\Delta B =1$ transitions at tree-level is (with $d_i=d{\rm ~or~}s$)
\begin{equation}
{\cal H} = \frac{G_F}{\sqrt{2}}\left[V_{cb}^\star V_{ud_i}
\left(C_1Q^n_1 +C_2  Q^n_2 \right)+
V_{ub}^\star V_{cd_i}
\left(C_1Q^c_1 +C_2  Q^c_2 \right)\right]+{\rm ~h.~c.}
\label{heffq}
\end{equation}
where the neutral modes ${B}^0\to \bar{D}^{0} (\bar{D}^{0\star})$ are mediated by $Q_1^n=(\bar{b} d_i)_{V-A}(\bar{u}c)_{V-A}$ and $Q_2^n=(\bar{b}c)_{V-A}(\bar{u}d_i)_{V-A}$; and the charged modes $B^+\to D^+ (D^{+\star})$  are mediated by $Q_1^c=(\bar{b}d_i)_{V-A}(\bar{c}u)_{V-A}$ and $Q_2^c=(\bar{b}u)_{V-A}(\bar{c}d_i)_{V-A}$ respectively \footnote{For our numerical estimates we will use the tree-level coefficients $C_1=0$, $C_2=1$. QCD corrections are known, but this difference is much smaller than other uncertainties in our calculation.}.
Our first task is to write the operators corresponding to 
Eq.~(\ref{heffq}) in the heavy quark effective theory.

For the charged modes this was already done in Ref.~\cite{Grinstein:1999qr}. The operators $Q^c_{1,2}$ can be written in terms of their heavy ($A,B$) and light ($a,b$) degrees of freedom in the more general form
\begin{equation}
{\cal O}^{ab \bar{A} B} = \bar{A}\gamma_\mu(1-\gamma_5)a\bar{B}\gamma^\mu(1-\gamma_5)b.
\label{chop}
\end{equation}
As pointed out by Grinstein and Lebed \cite{Grinstein:1999qr}, this form illustrates that there are symmetry relations that would allow one to extract the coupling of this effective Lagrangian from the measurements of $B-\bar{B}$ mixing.

For the charged transitions this operator has to 
destroy a heavy and light quarks of flavor $(\bar{A},a)=(\bar{b},u)$ and create a  heavy and light quarks of flavor $(\bar{B},b)=(c,\bar{d}_i)$. This operator transforms as a $(6_L,1_R)$ under the chiral symmetry and, as shown in Ref.~\cite{Grinstein:1992qt}, Eq.~(\ref{chop}) matches in the symmetry limit of the effective theory onto 
\begin{eqnarray}
{\cal O}^{ab \bar{A} B}& =&\beta_W  Tr \left[(\xi_{ac} \hqb{{A}}{c})\gamma_\mu (1-\gamma_5)\right]Tr\left[ (\xi_{bd} \hbq{B}{d})\gamma^\mu (1-\gamma_5) \right] 
\label{weakop}
\end{eqnarray}

As pointed out in Ref.~\cite{Grinstein:1999qr} this same operator with 
$(\bar{A},a)=(\bar{b},d)$ and $(\bar{B},b)=(b,\bar{d})$ is responsible for $B-\bar{B}$ mixing so that in principle the coefficient $\beta_W$ can be extracted from experiment (for this case there is an additional color factor of 8/3). 
This heavy quark symmetry relation is valid  in the GL limit where $m_B-m_D <<\Lambda_{QCD}$ and the four velocity of the $B$ and $D^\star$ mesons is the same. For our numerical estimates we will use
\begin{equation}
\beta_W= \frac{1}{4} f_Bf_{D}\sqrt{m_Bm_D} \sim (0.034\pm 0.009) {\rm ~GeV}^3,
\label{betaw}
\end{equation}
where we used the decay constants $f_B=(191\pm 27)~MeV$ \cite{Charles:2004jd} and $f_D =(225^{+11}_{-13}\pm 21)~MeV$ \cite{Simone:2004fr}. This last one is in good agreement with the recent CLEO measurement $f_D=(222.6\pm 16.7^{+2.8}_{-3.4}~MeV$ \cite{Artuso:2005ym}.

For the radiative decay modes that we consider in this paper there are no light pseudoscalars involved so we set $\xi=1$ in Eq.~(\ref{weakop}) 
to obtain the matrix elements:
\begin{eqnarray}
<D_v^*(\eta)|{\cal O}^{ab\bar{A}B}| B_v^*(\eta^{'})> &=& 4\beta_W   \eta \cdot \eta^{\prime\star}, \nonumber \\
<D_v |{\cal O}^{ab\bar{A}B}| B_v>& = &4\beta_W , \nonumber \\
<D_v^*(\eta) |{\cal O}^{ab\bar{A}B}| B_v>& = &-4\beta_W  \eta \cdot v, \nonumber \\
<B_v^*(\eta) |{\cal O}^{ab\bar{A}B}| D_v>& = &-4\beta_W  \eta \cdot v, 
\label{vertices}
\end{eqnarray}
where $\eta,\eta^\prime$ are the vector meson polarization vectors and $v$ the $B$ meson velocity. The last two terms vanish when the condition $v\cdot \eta =0 $ is used, we retain them because we will use them beyond leading order in the heavy quark expansion later on, as they contribute known terms of order $(m_b-m_c)/m_c$. 

The neutral modes involving the weak transition $B^0_i \rightarrow \bar{D}^0$ are mediated by the operators $Q^n_{1,2}$ in Eq.~(\ref{heffq}). To construct a matching operator in the effective theory we notice that they transform as $(8_L,1_R)$ under the chiral symmetry. We also need to extract the part of the operators responsible for destroying a heavy anti-quark (of flavor ${b}$) and creating a heavy anti-quark (of flavor $c$).  A possible match for $Q^n_2$ of the current-current form is
\begin{equation}
{\cal O } = Tr \left[ \hbq{\bar{c}}{a}\gamma_\mu (1-\gamma_5) \hq{\bar{b}}{f} \right] \left( \xi^\dagger \partial^\mu \xi \right)_{fa}.
\end{equation}
However, this operator does not contribute to the processes without light pseudo-scalars that we are discussing.

A possible operator that does contributes to $B^0_i \rightarrow \bar{D}^0$ transitions at tree level is of the form 
\begin{eqnarray}
{\cal O} &=& \beta_W^\prime Tr\left[ (\hbq{\bar{c}}{j}\xi_{ju}^{\dagger})\gamma_\mu(1-\gamma_5)\right]Tr\left[ (\xi_{dk}\hq{\bar{b}}{k})\gamma^\mu(1-\gamma_5)\right].
\end{eqnarray}
This operator leads to matrix elements analogous to those in Eq.(\ref{vertices}) with $\beta_W \to \beta_W^\prime$. As was the case with the charged modes, there are other possible matches but they lead to the same result \cite{Grinstein:1992qt}.  We have not found a way to determine $\beta_W^\prime$ from symmetry relations and must resort to the factorization model,
\begin{equation}
\beta_W^\prime =\frac{1}{12}f_B f_D \sqrt{m_B m_D}=\frac{\beta_W}{N_c}.
\label{betawp} 
\end{equation}
An additional factor of $1/N_c=1/3$ relative to $\beta_W$ occurs because the contribution of $Q^n_{2}$ to these weak transitions in factorization requires a Fierz transformation and color rearrangement.

For weak transitions involving the positive parity states,  $B\to D^{\star\star}$, we have two new operators for the $S$ doublet. 
The  operators $Q_2^c$ and $Q_2^n$ 
in Eq.~(\ref{heffq}) for the charged and neutral modes can be matched in the factorization approximation into the operators
\begin{eqnarray}
Q_2^c &\to&- \beta_W Tr\left[ (\hbq{\bar{c}}{j}\xi_{ju}^{\dagger})\gamma_\mu(1-\gamma_5)\right]Tr\left[ \gamma^\mu(1-\gamma_5)(\xi_{dj}S^{(\bar{b})}_j)\right] 
 \nonumber \\
Q_2^n &\to& -\beta_W^\prime Tr\left[ (\hbq{\bar{c}}{j}\xi_{ju}^{\dagger})\gamma_\mu(1-\gamma_5)\right]Tr\left[ \gamma^\mu(1-\gamma_5)(\xi_{dj}S^{(\bar{b})}_j)\right] 
\label{weaks}
\end{eqnarray}
for transitions of the form $(0^-,1^-)_b \to (0^+,1^+)_c$. The extra minus sign is chosen so that the coefficients $\beta_W$ and $\beta_W^\prime$ are the same as those in Eqs.~(\ref{betaw})~and~(\ref{betawp}) if we take the decay constants and masses of the two doublets to be the same. This is approximately true for the decay constants in the analysis of Ref.~\cite{Colangelo:1992kc} based on QCD sum rules.  

The $T=(1^+,2^+)$ multiplet does not participate in the weak transitions in this approximation since its decay constant vanishes \cite{Casalbuoni:1992dx}.

The matrix elements obtained from Eq.~(\ref{weaks}) (and corresponding ones  for $(0^-,1^-)_b\to (0^+,1^+)_c$ transitions) that do not involve light mesons are then
\begin{eqnarray}
<D_v^*(\eta)|Q_2^c| B_{1v}(\eta^{'})> &=& -4\beta_W   \eta \cdot \eta^{'}, \nonumber \\
<D_v |Q_2^c| B_{0v}>& = &-4\beta_W  , \nonumber \\
<D_v^*(\eta) |Q_2^c| B_{0v}>& = &4\beta_W  \eta \cdot v, \nonumber \\
<D_v|Q_2^c|B_{1v}(\eta^\prime)>&=& 4\beta_W  \eta^\star \cdot v,
\label{verticesst}
\end{eqnarray}
and the same expressions with $\beta_W\to \beta_W^\prime$ for the neutral modes. Finally we will need weak transitions from $(0^+,1^+)_b\to (0^+,1^+)_c$. They follow from an effective Lagrangian like the one in Eq.~(\ref{weakop}) with $S$ fields replacing $H$ fields and produce matrix elements with the same sign as those in Eq.~(\ref{vertices}).

\section{$B \rightarrow D^\star \gamma$}

\begin{figure}[htb]
\includegraphics[width=4 in]{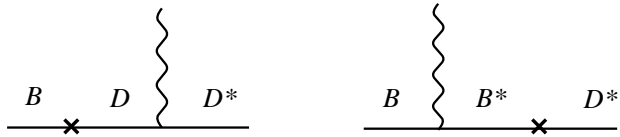}
\caption{Pole diagrams responsible for $B \to D^\star \gamma$ at leading order in heavy quark and chiral theories.}
\label{f:bdgdiags}
\end{figure}

We now turn our attention to the single radiative decay for exclusive channels. Both the charged $B^+ \to D^{\star +} \gamma$ and neutral
${B}^0 \to \bar{D}^{\star 0} \gamma$ modes have been estimated before. We begin with the charged mode  discussion of Ref.~\cite{Grinstein:1999qr}. With the ingredients introduced in the previous section it is straight-forward to compute the amplitude from the two diagrams in Figure~\ref{f:bdgdiags}.
For the process $B_v^+\to D_{v i}^{\star +}(\eta) \gamma(q,\epsilon)$ we find in terms of the notation defined in the appendix, a leading order amplitude containing only a magnetic form factor: 
$F_M= (F_M)_{LO} r^+_\mu $ with,
\begin{eqnarray}
\left(F_M\right)_{LO}(B^+ \to D^{\star +}_i\gamma) &\equiv& \sqrt{2}G_F V_{ub}^\star V_{cd_i} \beta_W e\left(\frac{\mu_{D^+}}{m_{B^+}-m_{D^{\star+}}}\right) \nonumber \\
r_\mu^+ &\equiv& \left(1+\frac{\mu_{B^+}}{\mu_{D^+}} \frac{2m_{B^+}}{m_{B^+}+m_{D^{\star+}}}\right).
\label{gamp}
\end{eqnarray}
Similarly,  for the neutral modes $F_M = (F_M)_{LO} r^0_\mu $ with 
\begin{eqnarray}
(F_M)_{LO}(B^0_i \to \bar{D}^{\star 0}\gamma) &=& \sqrt{2}G_F V_{cb}^\star V_{ud_i} \beta_W^\prime  e\left(\frac{\mu_{D^0}}{m_{B^0}-m_{D^{\star 0}}}\right)\nonumber \\
r_\mu^0 &\equiv& \left(1+\frac{\mu_{B^0}}{\mu_{D^0}} \frac{2m_{B^0}}{m_{B^0}+m_{D^{\star0}}}\right).
\label{gampn}
\end{eqnarray}

Using the leading order magnetic moments  and the GL limit to evaluate these amplitudes yields $r^{+}_\mu = -1$, $r^0_\mu=1/2$. Combining this 
with exact kinematics for the phase space one obtains \footnote{Our rate for the charged process is a factor of 9 smaller than the result obtained in Ref.~\cite{Grinstein:1999qr} which corresponds to this limit. It appears that Ref.~\cite{Grinstein:1999qr} incorrectly used the light antiquark charge to calculate the magnetic moments instead of the light quark charge. }
\begin{eqnarray}
\Gamma(B^+\to D^{\star +}_i\gamma) &=& \frac{G_F^2}{36}\left|V_{ub}^\star V_{cd_i}\right|^2\alpha \left({\beta_W\beta r^+_\mu}\right)^2\frac{m_D^\star}{m_B^4} (m_B-m_{D^\star})(m_B+m_{D^\star})^3 \nonumber \\
\Gamma(B^0_i\to \bar{D}^{\star 0}\gamma) &=& \frac{G_F^2}{9}\left|V_{cb}^\star V_{ud_i}\right|^2\alpha \left({\beta_W^\prime\beta r^0_\mu}\right)^2\frac{m_D^\star}{m_B^4} (m_B-m_{D^\star})(m_B+m_{D^\star})^3.
\label{grate}
\end{eqnarray}

We wish to emphasize that these results exhibit a high sensitivity to non-leading contributions to the magnetic moments.   With exact kinematics (i.e. taking $m_B \neq m_D$) and using the three models for magnetic moments in Table~8 of Ref.~\cite{Casalbuoni:1996pg} one finds 
\begin{eqnarray}
\chi-{\rm loop}:&&  r^+_\mu = -6.3, \ \ r^0_\mu = 0.28,
\nonumber \\
VMD:&&  r^+_\mu = -4.7, \ \ r^0_\mu = 0.38,
\nonumber \\
RQM:&&  r^+_\mu = -4.7, \ \ r^0_\mu = 0.50.
\end{eqnarray}
These numbers lead to predicted rates that are larger than the leading order prediction by factors between 16 and 36 for the charged modes. For the neutral modes the effect is more modest. Allowing the magnetic moments to vary in the ranges predicted by these models results in the rates
\begin{eqnarray}
0.7\times 10^{-7}\lsim&B({B}^0\to \bar{D}^{\star 0}\gamma)&\lsim 4.6\times 10^{-7} ,         \nonumber \\
1.5\times 10^{-8}\lsim&B({B}^0_s\to \bar{D}^{\star 0}\gamma)&\lsim   3.0\times 10^{-8} ,    \nonumber \\
3.6\times 10^{-9}\lsim&B(B^+\to D^{\star +}\gamma)&\lsim     4.9\times 10^{-9},         \nonumber \\
0.7\times 10^{-7}\lsim&B(B^+\to D^{\star +}_s\gamma)&\lsim    1.0\times 10^{-7}.   
\label{bdgrangelo}     
\end{eqnarray}
There is also a significant uncertainty from the use of $g_M(0)$ for the electromagnetic transitions, when in these reactions one should use  $g_M(k^2\sim-E_\gamma^2\sim-(2~{\rm GeV})^2)$. 

\begin{figure}[htb]
\includegraphics[width=6 in]{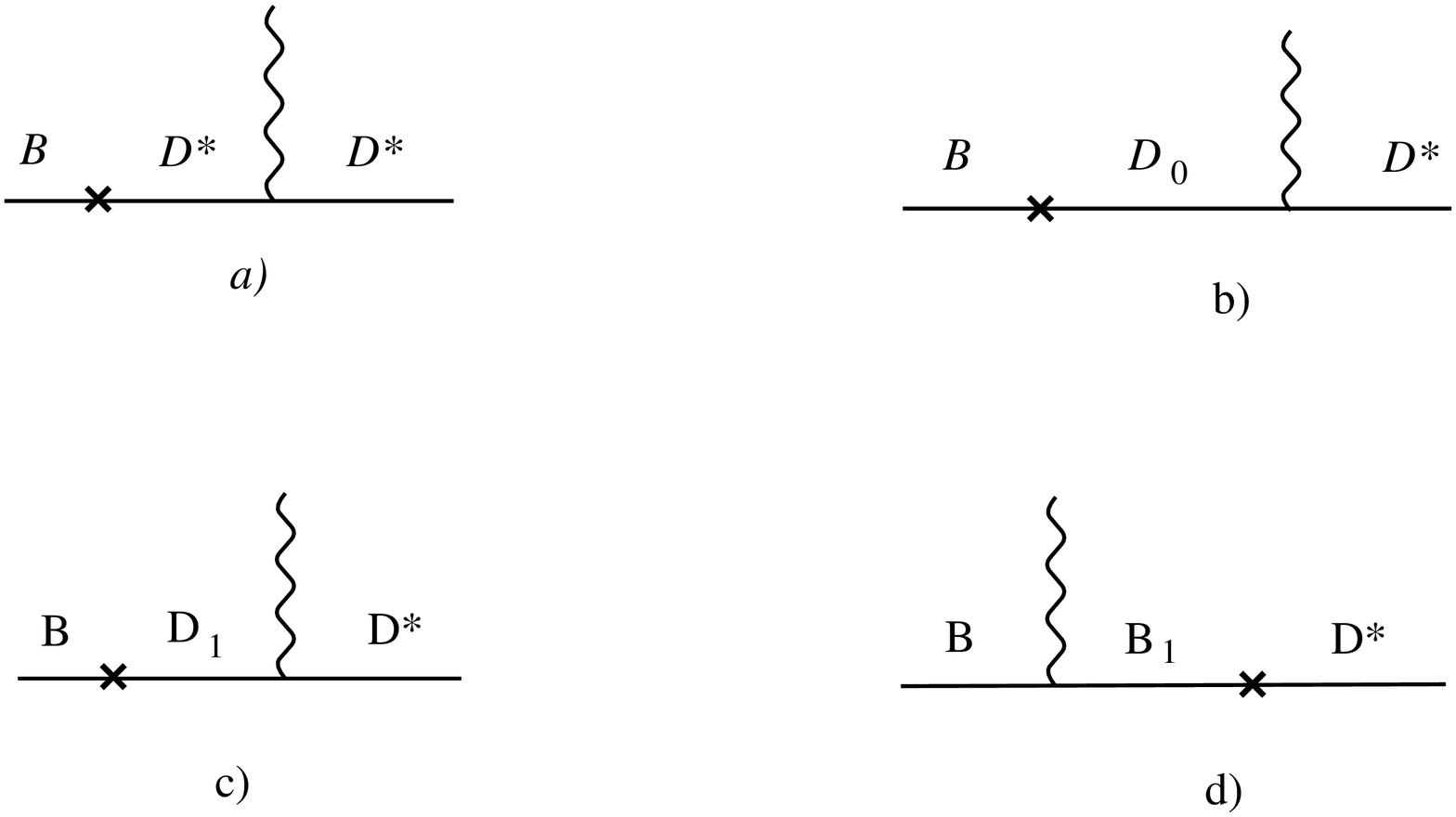}
\caption{Non-leading contributions to $B\to D^\star \gamma$: a)$P^*-P$ weak transitions that vanish in the $m_B = m_D$ limit; b-d) additional pole contributions from positive parity states.}
\label{f:extrad}
\end{figure}

We now consider two types of higher order corrections to these results that are counterpart to additional terms present in the pole model calculation of the neutral modes in Ref.~\cite{Cheng:1992xi,Cheng:1994kp}. They correspond to the diagrams shown in Figure~\ref{f:extrad}.
The first diagram (Figure~\ref{f:extrad}a) contributes beyond leading order in the heavy quark symmetry when we allow for $m_B\neq m_D$. It generates an electric amplitude from corrections to the leading order $1^-$ propagator fixed by reparametrization invariance
\begin{equation}
F_E(B^+ \to D^{\star +}_i\gamma) = -\sqrt{2}G_F V_{ub}^\star V_{cd_i} \beta_W e\left(\frac{\mu_{D^{\star+}}}{m_{B^+}-m_{D^{\star+}}}\right)\left(1-\frac{m_B^2}{m_D^2}\right).
\end{equation}
The next two diagrams (Figure~\ref{f:extrad}b,c) involve an intermediate positive parity state that can be either the $D_0$ or the $D_1$ from the $S$ multiplet. They generate an electric and magnetic form factor respectively that can be written as
\begin{eqnarray}
F_E(B^+ \to D^{\star +}_i\gamma) &=&- \sqrt{2}G_F V_{ub}^\star V_{cd_i} \beta_W e\left(\frac{\mu^S_{D^+}}{m_{B^+}-m_{D_0}}\right) \nonumber \\
F_M(B^+ \to D^{\star +}_i\gamma) &=&-\sqrt{2}G_F V_{ub}^\star V_{cd_i} \beta_W e\left(\frac{\mu^S_{D^+}}{m_{B^+}-m_{D_1}}\right)\left(1-\frac{m_B^2}{m_{D_1}^2}\right).
\end{eqnarray}
Finally the last diagram (Figure~\ref{f:extrad}d) involving an intermediate $B_1$ meson contributes to the electric form factor as
\begin{equation}
F_E(B^+ \to D^{\star +}_i\gamma) = -\sqrt{2}G_F V_{ub}^\star V_{cd_i} \beta_W e\mu^S_{B^+}\left(\frac{2m_B}{m_B^2-m_{D^\star}^2}\right).
\end{equation}
Analogous results are obtained for the neutral mode with the obvious replacement $V_{ub}^\star V_{cd_i}\beta_W\to V_{cb}^\star V_{ud_i}\beta_W^\prime$ and magnetic moments and masses appropriate for neutral mesons. Combining all these partial results we finally obtain
\begin{eqnarray}
F_M(B^+\to D^{\star +}_i\gamma) &=& \sqrt{2}G_F V_{ub}^\star V_{cd_i} \beta_W  e \left(\frac{\mu_{D^+}}{m_{B^+}-m_{D^{\star+}}}\right)
 \\
&\cdot&
 \left(1+\frac{\mu_{B^+}}{\mu_{D^+}} \frac{2m_{B^+}}{m_{B^+}+m_{D^{\star+}}}+
\frac{\mu^S_{D^+}}{\mu_{D^+}} \frac{m_B^2-m_{D_1^+}^2}{m_{D^+_1}^2}  \frac{m_B-m_{D^{*+}}}{m_B-m_{D_1^+}}\right), \nonumber \\
F_E(B^+\to D^{\star +}_i\gamma) &=& -\sqrt{2}G_F V_{ub}^\star V_{cd_i} \beta_W  e\left(\frac{\mu^S_{D^+}}{m_{B^+}-m_{D^{*+}}}\right)\nonumber \\
&\cdot&
 \left(\frac{m_B-m_{D^{*+}}}{m_B-m_{D_0^+}}-\frac{\mu_{D^{*+}}}{\mu^S_{D^+}} \frac{m_B^2-m_{D^{*+}}^2}{m_{D^{*+}}^2}+
\frac{\mu^S_{B^+}}{\mu^S_{D^+}}\frac{2m_B}{m_B+m_{D^{*+}}}\right). \nonumber
\end{eqnarray}
With the range of magnetic moments provided by the three models, we find 
\begin{eqnarray}
1.2\times 10^{-5}\lsim&B({B}^0\to \bar{D}^{\star 0}\gamma)&\lsim 3.1\times 10^{-5} ,         \nonumber \\
0.7\times 10^{-6}\lsim&B({B}^0_s\to \bar{D}^{\star 0}\gamma)&\lsim   1.7\times 10^{-6} ,    \nonumber \\
0.6\times 10^{-7}\lsim&B(B^+\to D^{\star +}\gamma)&\lsim     1.0\times 10^{-7},         \nonumber \\
0.6\times 10^{-6}\lsim&B(B^+\to D^{\star +}_s\gamma)&\lsim    1.4\times 10^{-6}.   
\label{bdgrange}     
\end{eqnarray}
These ranges indicate only the uncertainty in the magnetic moments, in particular they do not include the uncertainty in $\beta_W$ or any other parameters. For comparison, Grinstein and Lebed obtained 
$B(B^+\to D^{\star +}_s\gamma)=2 \times 10^{-8}$ \cite{Grinstein:1999qr} (when we correct their number for the missing factor 1/9). As mentioned before, the charge mode exhibits a large sensitivity to the value of the magnetic moments due to a partial cancellation between the two terms in Eq.~(\ref{gamp}). This sensitivity is milder when the additional terms are included as can be seen from the range in Eq.~(\ref{bdgrange}). Similarly we can compare our result to that of Ref.~\cite{Cheng:1994kp}, $B(\bar{B}^0 \to D^{\star 0} \gamma) = 9.2 \times 10^{-7}$. Again this result has the same order of magnitude as the leading order contribution, Eq.~(\ref{gampn}), and the larger number in Eq.~(\ref{bdgrange}) is due to the contributions of the positive parity states. The prediction for $B(\bar{B}^0 \to D^{\star 0} \gamma)$ in Eq.~(\ref{bdgrange}) is in fact close to the experimental upper bound ${\cal B}(\bar{B}^0 \to D^{\star 0} \gamma)< 2.5\times 10^{-5}$ \cite{Aubert:2005aj} and part of the range is already excluded. For comparison, a recent calculation finds ${\cal B}(\bar{B}^0 \to D^{\star 0} \gamma)\sim 1.6 \times 10^{-6}$ \cite{MacdonaldSorensen:2006ds}
using a different framework. Although smaller than our range in Eq.~(\ref{bdgrange}), this result is not incompatible with ours given the large uncertainty illustrated by the differences between Eqs.~(\ref{bdgrangelo})~and~(\ref{bdgrange}).

\section{$B \rightarrow D \gamma \gamma $ and HQET}
 
\begin{figure}[htb]
\includegraphics[width=5 in]{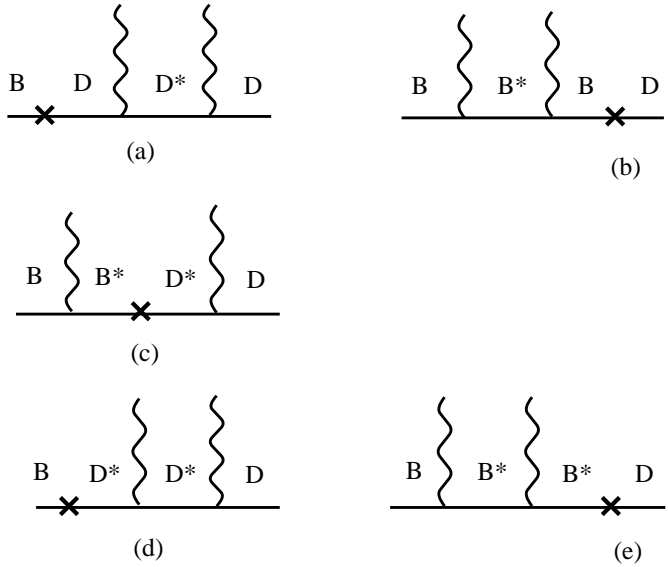}
\caption{Pole diagrams responsible for $B \to D^\star \gamma \gamma$ at leading order in heavy quark and chiral theories.}
\label{diags}
\end{figure}

We are now in a position to estimate the $B \to D \gamma \gamma$ amplitudes. For the process $B \rightarrow D \gamma \gamma $ there are 5  diagrams involving only the $H$ doublet,  shown schematically on Fig.\ref{diags}. Diagrams Fig.\ref{diags} (d) and (e) vanish at leading order due to the condition $v_{\mu}(g^{\mu\nu}-v^{\mu}v^{\nu})=0$ at the weak vertex and we left with three LO diagrams.
Two of these diagrams have a $B-D$ weak transition and the third one 
a  $B^\star-D^\star$ weak transition. In all cases the two photons are emitted from magnetic dipole couplings on the external legs.

A straightforward calculation then yields the desired amplitude for $B^0 \rightarrow \bar{D}^0 \gamma \gamma $. In terms of the form factors defined in the Appendix we find
\begin{eqnarray}
A&=&- C_n
\left[\frac{\mu_D^2}{\Delta m}(\frac{1}{\Delta m -E_1} + \frac{1}{\Delta m- E_2} ) + \frac{\mu_B^2}{E_1 E_2} \right.\nonumber \\
&+&\left. \mu_B \mu_D (\frac{1}{E_1(\Delta m-E_1)} +\frac{1}{E_2(\Delta m-E_2)}  )\right], \nonumber \\
B&=&-\frac{A}{2},
\end{eqnarray}
where $E_1$ and $E_2$ are photons energies, $\Delta m \equiv m_B-m_D$, and we have defined
\begin{equation}
C_n \equiv  2\sqrt{2} \pi G_F \alpha_{em} \beta_W^\prime V^\star_{cb} V_{ud} M_B^2   
\end{equation}

\begin{figure}[htb]
\includegraphics[width=5in]{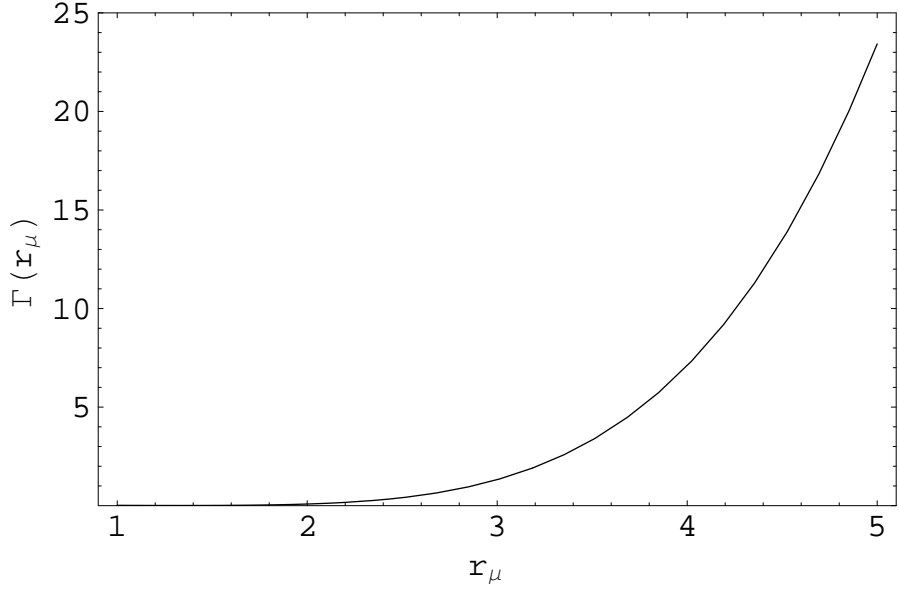}
\caption{$\Gamma(B^0\to \bar{D}^0 \gamma \gamma)$  as a function of $r_\mu$. We use the RQM magnetic moment for the $B$ and we normalize the rate to its value when $r_\mu \sim 2.87$,  the value of $\mu_D$ in the RQM. }
\label{bdggran}
\end{figure}

Using these lowest order form factors we find a large range for the double radiative decay rates depending on the model used for the magnetic moments, for the mode with most favored CKM angles,
\begin{equation}
0.3\times 10^{-10} \lsim B(B^0\to \bar{D}^0 \gamma \gamma) \lsim 3.5\times 10^{-10}.
\end{equation}
This large sensitivity to the input parameters is due in part to a cancellation between the terms involving the magnetic moments for the charm and bottom mesons. To illustrate this we write 
\begin{equation}
\mu_D = -r_\mu \mu_B,
\end{equation}
and show in Figure~\ref{bdggran}  $\Gamma(B^0\to \bar{D}^0 \gamma \gamma)$  as a function of $r_\mu$. We use the RQM magnetic moment for the $B$ and we normalize the rate to its value when $r_\mu \sim 2.87$,  the value of $\mu_D$ in the RQM. 

\begin{figure}[htb]
\includegraphics[width=5in]{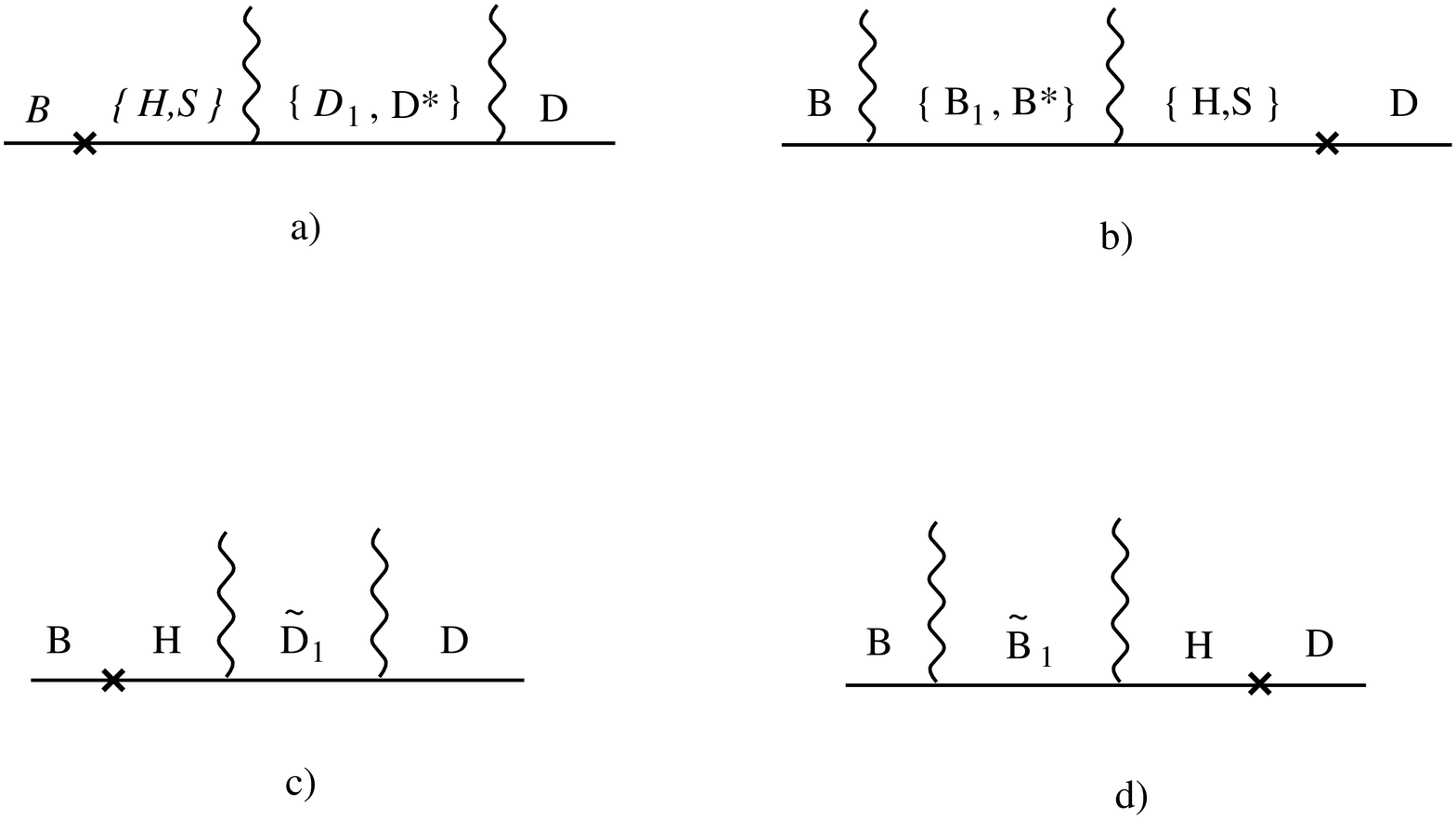}
\caption{Additional diagrams involving particles from the $(0^+,1^+)$ and $(1^+,2^+)$ doublets. H(S) stands for either of the H(S)-multiplet members}
\label{bdggnlo1}
\end{figure}

\begin{figure}[htb]
\includegraphics[width=5in]{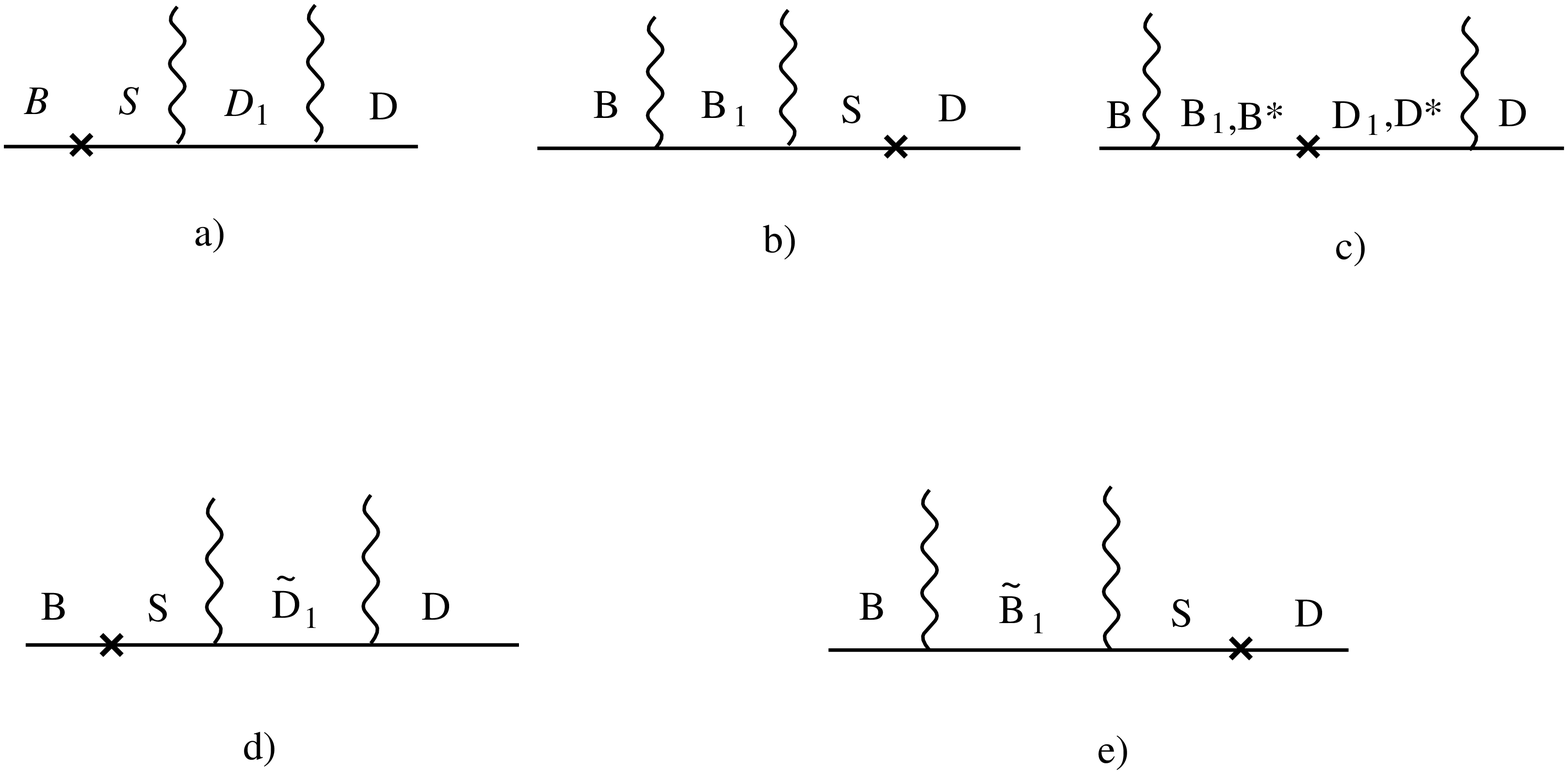}
\caption{Additional diagrams involving particles from the $(0^+,1^+)$ and$(1^+,2^+)$ doublets. S stands for either of the S-multiplet members}
\label{bdggnlo2}
\end{figure}

We next consider two other types of contribution. First the non-leading contributions from diagrams d and e. They arise from corrections to the $1^-$ propagator that are fixed by reparametrization invariance resulting in
\begin{eqnarray}
C &=&\frac{C_n}{2}\left[\frac{\mu_D\mu_{D^\star}}{\Delta m}\frac{m_B^2-m_{D}^2}{m_{D}^2}\left(\frac{1}{\Delta m-E_1}+\frac{1}{\Delta m-E_2}\right)+\frac{\mu_B\mu_{B^\star}}{E_1E_2}\frac{m_{B}^2-m_{D}^2}{m_{B}^2}\right]  \\
D &=&-\frac{C_n}{2}\left[\frac{\mu_D\mu_{D^\star}}{\Delta m}\frac{m_B^2-m_{D}^2}{m_{D}^2}\left(\frac{1}{\Delta m-E_1}-\frac{1}{\Delta m-E_2}\right)-\frac{\mu_B\mu_{B^\star}(E_2-E_1)}{E_1E_2(E_1+E_2)}\frac{m_{B}^2-m_{D}^2}{m_{B}^2}\right] \nonumber.
\end{eqnarray}
Additional contributions arise when the positive parity states appear as in Figure~\ref{bdggnlo1}. The contributions from the $S$ multiplet are
\begin{eqnarray}
A&=&-C_n\left[ \frac{\mu^S_D\mu_{D}}{\Delta m}\frac{m_B^2-m_D^2}{m_D^2}\left(\frac{1}{\Delta m-E_1}+\frac{1}{\Delta m-E_2}\right)-\frac{\mu^S_B\mu_{B}}{E_1E_2}\frac{m_B^2-m_D^2}{m_B^2}\right]  \\
B&=&\frac{C_n}{2}\left[ \frac{\mu^S_D(\mu^S_D+\mu_{D}\frac{m_B^2-m_D^2}{m_D^2})}{\Delta m}\left(\frac{1}{\Delta m-E_1}+\frac{1}{\Delta m-E_2}\right)+\frac{\mu^S_B(\mu^S_B-\mu_{B}\frac{m_B^2-m_D^2}{m_B^2})}{E_1E_2}\right] \nonumber \\
C&=&-\frac{C_n}{2}\left[\frac{\mu^S_D}{\Delta m}
(\mu_D-\mu^S_D\frac{m_B^2-m_D^2}{m_D^2})\left(\frac{1}{\Delta m-E_1}+\frac{1}{\Delta m-E_2}\right)+\frac{\mu^S_B}{E_1E_2}(\mu_B-\mu^S_B\frac{m_B^2-m_D^2}{m_B^2})\right]\nonumber \\
D&=&\frac{C_n}{2}[\hspace{3mm}\frac{\mu^S_D}{\Delta m}
(\mu_D+\mu^S_D\frac{m_B^2-m_D^2}{m_D^2})\left(\frac{1}{\Delta m-E_1}-\frac{1}{\Delta m-E_2}\right) \nonumber \\
&-&\frac{\mu^S_B(E_2-E_1)}{E_1E_2(E_1+E_2)}(\mu_B+\mu^S_B\frac{m_B^2-m_D^2}{m_B^2})\hspace{3mm}] \nonumber
\end{eqnarray}
The contributions from intermediate $T$ doublet states are
\begin{eqnarray}
B&=&4C_n\left[ \frac{(\mu^T_D)^2}{\Delta m}\left(\frac{1}{\Delta m-E_1}+\frac{1}{\Delta m-E_2}\right)+\frac{(\mu^T_B)^2}{E_1E_2}\right]  \\
C&=&-2C_n\left[\frac{(\mu^T_D)^2}{\Delta m}
\frac{m_{B}^2-m_{D}^2}{m_{D}^2}\left(\frac{1}{\Delta m-E_1}+\frac{1}{\Delta m-E_2}\right)+\frac{(\mu^T_B)^2}{E_1E_2}\frac{m_{B}^2-m_{D}^2}{m_{B}^2}\right]\nonumber \\
D&=&-2C_n\left[\frac{(\mu^T_D)^2}{\Delta m}
\frac{m_{B}^2-m_{D}^2}{m_{D}^2}\left(\frac{1}{\Delta m-E_1}-\frac{1}{\Delta m-E_2}\right)-\frac{(\mu^T_B)^2(E_2-E_1)}{E_1E_2(E_1+E_2)}\frac{m_{B}^2-m_{D}^2}{m_{B}^2}\right] \nonumber
\end{eqnarray}

Additional contributions from SS and TS multiplet transitions as in Figure~\ref{bdggnlo2}. We split them into those from diagrams with two members of the $S$ doublet:
\begin{eqnarray}
B&=& \frac{C_n}{2}
\left[\frac{\mu^S_D \mu_{D^\star}\frac{m_B^2-m_D^2}{m_D^2}}{\Delta m}(\frac{1}{\Delta m -E_1} + \frac{1}{\Delta m- E_2} ) - \frac{\mu^S_B\mu_{B^\star}\frac{m_B^2-m_D^2}{m_B^2}}{E_1 E_2} \right.\nonumber \\
&+&\left. \mu^S_B \mu^S_D (\frac{1}{E_1(\Delta m-E_1)} +\frac{1}{E_2(\Delta m-E_2)}  )\right], \nonumber \\
C&=&-\frac{C_n}{2}[ \frac{\mu^S_D\mu_{D}}{\Delta m}\left(\frac{1}{\Delta m-E_1}+\frac{1}{\Delta m-E_2}\right)+\frac{\mu^S_B\mu_{B}}{E_1E_2} \nonumber \\
&-&(\mu^S_B \mu_{D}-\mu^S_D \mu_{B}) (\frac{1}{E_1(\Delta m-E_1)} +\frac{1}{E_2(\Delta m-E_2)}  )],\\
D&=&-\frac{C_n}{2}[ \frac{\mu^S_D\mu_{D}}{\Delta m}\left(\frac{1}{\Delta m-E_1}-\frac{1}{\Delta m-E_2}\right)-\frac{\mu^S_B\mu_{B}(E_2-E_1)}{E_1E_2(E_1+E_2)}\nonumber \\
&+&(\mu^S_B \mu_{D}+\mu^S_D \mu_{B}) (\frac{1}{E_1(\Delta m-E_1)} -\frac{1}{E_2(\Delta m-E_2)}  )] ,\nonumber
\end{eqnarray}
and those from diagrams with one member of the $S$ doublet and one member of the $T$ doublet:
\begin{eqnarray}
B&=&\frac{2C_n}{\sqrt{3}}\left[ \frac{\mu^T_D\mu^{TS}_D}{\Delta m}\frac{m_B^2-m_D^2}{m_D^2}\left(\frac{1}{\Delta m-E_1}+\frac{1}{\Delta m-E_2}\right)+\frac{\mu^T_B\mu^{TS}_B}{E_1E_2}\frac{m_B^2-m_D^2}{m_B^2}\right]  \nonumber \\
C&=&\frac{4C_n}{\sqrt{3}}\left[\frac{\mu^T_D\mu^{TS}_D}{\Delta m}
\left(\frac{1}{\Delta m-E_1}+\frac{1}{\Delta m-E_2}\right)-\frac{\mu^T_B\mu^{TS}_B}{E_1E_2}\right]  \\
D&=&\frac{4C_n}{\sqrt{3}}\left[\frac{\mu^T_D\mu^{TS}_D}{\Delta m}
\left(\frac{1}{\Delta m-E_1}-\frac{1}{\Delta m-E_2}\right)+\frac{\mu^T_B\mu^{TS}_B(E_2-E_1)}{E_1E_2(E_1+E_2)}\right] .\nonumber
\end{eqnarray}

In a similar manner we obtain the result for  $B^{\pm} \rightarrow D^{\pm} \gamma \gamma $ with the replacements $V_{cb}V_{ud}\beta_W^\prime \to V_{ub}V_{cd} \beta_W$, for $B^0_s \rightarrow D^0 \gamma \gamma$ with $V_{ud} \to V_{us}$
and $B^\pm \rightarrow D^\pm_s \gamma \gamma $ with $V_{cb}V_{ud}\beta_W^\prime \to V_{ub}V_{cs} \beta_W$.

Numerically we find the following ranges for the branching ratios when we vary the magnetic moments over the range predicted in the three different models
\begin{eqnarray}
1.7 \times 10^{-8} \lsim &B(B^0 \rightarrow \bar{D}^0 \gamma \gamma )& \lsim 8.0 \times 10^{-8} , \nonumber \\
0.8 \times 10^{-9}\lsim &B(B^0_s \rightarrow \bar{D}^0 \gamma \gamma) & \lsim 4.3 \times 10^{-9}, \nonumber \\
0.6 \cdot 10^{-11}\lsim &B(B^{\pm} \rightarrow D^{\pm} \gamma \gamma) & \lsim 2.0 \cdot 10^{-11}, \nonumber \\
1.1 \cdot 10^{-10}\lsim &B(B^{\pm} \rightarrow D^{\pm}_s \gamma \gamma) &\sim 3.6 \cdot 10^{-10} .
\label{bdggrange}
\end{eqnarray}
Once again these ranges include only variations of the rates with the magnetic moments in the $\chi$LM, VMD and RQM. They do not include other uncertainties such as that in the value of $\beta_W$.

\begin{figure}[htb]
\includegraphics[width=5.5in]{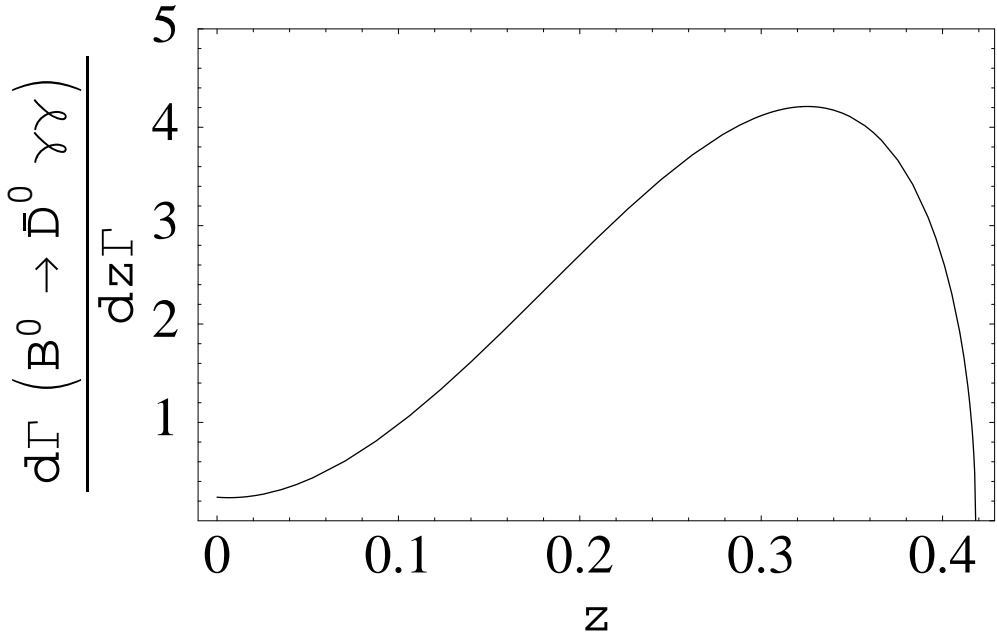}
\centering
\caption{Normalized differential decay rate for $B^0 \to \bar{D}^0 \gamma \gamma$ as a function of the photon pair invariant mass.}
\label{diffz}
\end{figure}

\begin{figure}[htb]
\includegraphics[width=4.5in]{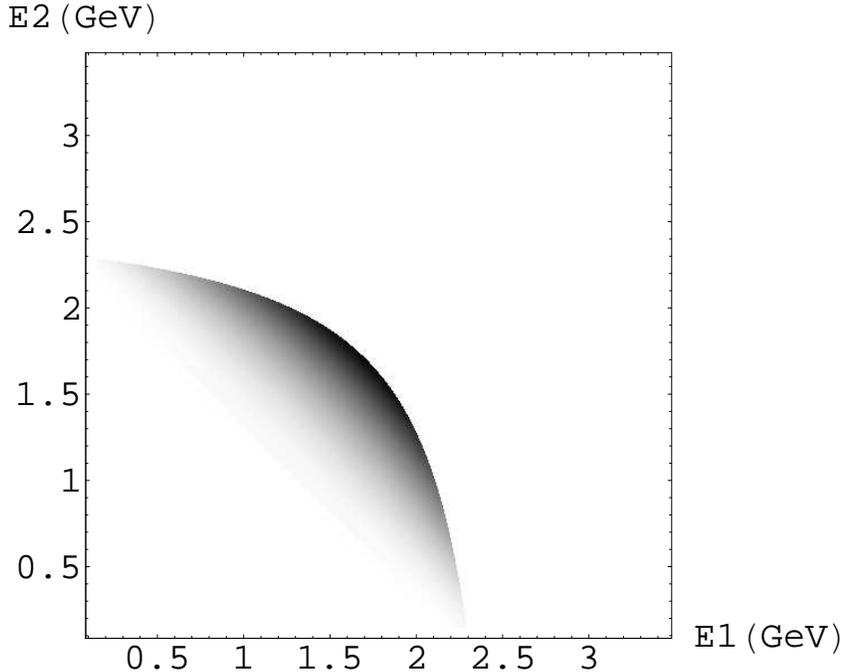}
\centering
\caption{Density plot for the double differential decay rate for $B^0 \to \bar{D}^0 \gamma \gamma$ as a function of the two photon energies.}
\label{densp}
\end{figure}

It is instructive to examine some of the features of the differential decay rates and we do so for the mode with most favorable CKM angles, $B^0 \to \bar{D}^0 \gamma \gamma$. We first plot in Figure~\ref{diffz} the normalized differential decay rate as a function of $z$, the dimensionless photon pair invariant mass defined in Eq.~(\ref{kindef}). The distribution does not have any thresholds as our calculation does not include absorptive parts. It is peaked at the higher invariant masses.
To evaluate the limits of the heavy quark expansion we first show in Figure~\ref{densp} the double differential decay rate as a function of the energies of the two photons as a density plot. The darker regions correspond to the most populated ones. The distribution is dominated by the region in which both photons tend to have similar energy between 1 and 2 GeV. This indicates that the approximation of constant magnetic moments for the photon emission vertices is slightly better for the double radiative decay modes than it was for the single radiative decay mode where $E_\gamma \sim 2.3$~GeV. However, it is clear that a substantial uncertainty remains due to this approximation. 

\begin{figure}[htb]
\includegraphics[width=4.5in]{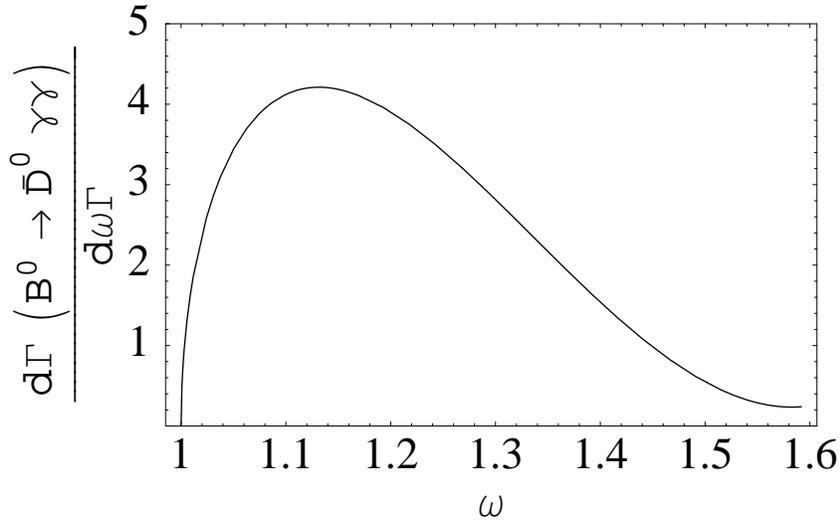}
\centering
\caption{Normalized differential decay rate for $B^0 \to \bar{D}^0 \gamma \gamma$ as a function of $\omega=v\cdot v^\prime$.}
\label{diffw}
\end{figure}

Finally we show again the normalized differential decay rate as a function of $\omega=v\cdot v^\prime$ in Figure~\ref{diffw}.  Here we see that the distribution is peaked at $\omega \sim1.1$, not too far from the symmetry limit. The heavy quark expansion should be better behaved for these modes than it is for the single radiative decays where $\omega \sim 1.5$.

\subsection{Additional Contributions}

We turn our attention to potential contributions with a different topology that have not appeared so far. Specifically we have in mind contributions in which the weak decay $B \to D X$ is followed by the electromagnetic $X\to \gamma\gamma$ vertex. We do not have a systematic way to include these contributions, but we illustrate them with a few examples. In any case, if there is a dominant contribution of this form, the photon pair invariant mass would be concentrated around $M_X$ and would be easy to isolate experimentally.

An example of this topology with a $c\bar{c}$ resonance  is $B \to D \eta_c \to D \gamma \gamma$. Using the narrow width approximation one finds $\Gamma(B \rightarrow D\gamma\gamma)=\Gamma(B \rightarrow \eta_c D) \cdot {\cal B}(\eta_c \rightarrow \gamma \gamma)$.
Experimentally it is known that  ${\cal B}(\eta_c \rightarrow \gamma \gamma)=4.3 \cdot 10^{-4}$ \cite{Eidelman:2004wy}. For the weak vertex we can use a recent pQCD calculation as an illustration, 
${\cal B}(B \rightarrow \eta_c D)=1.28 \cdot 10^{-5} $ \cite{Li:2005vr}.
This then leads to a contribution 
${\cal B}(B^0 \rightarrow {\bar D}^0\gamma\gamma)\sim 5.5 \cdot 10^{-9} $ smaller than our result in Eq.~\ref{bdggrange}. 
A different estimate ${\cal B}(B \rightarrow \eta_c D)=1.52 \cdot 10^{-7} $ \cite{Eilam:2001kw} would make this contribution even smaller. 
Another charmonium resonance with a measured two photon width is the $\chi_{c0}(1P) (0^+) $. Since the measured branching ratio is 
${\cal B}(\chi_{c0}(1P) (0^+) \rightarrow \gamma \gamma)=2.6 \cdot 10^{-4}$, similar to that for $\eta_c\to \gamma\gamma$ we expect at most a similar contribution to the double radiative $B$ decay.

We expect much larger contributions from the light pseudoscalars. This can be illustrated using the measured rates for ${\cal B}(B^0\to \bar{D}^0 X)$, $X=\pi^0,\eta,\eta^\prime$ which are all at the $10^{-4}$ level, and the respective two photon widths which range from nearly 100\% for the $\pi^0$ to a few percent for the $\eta^\prime$. These contributions would however, be easily separated experimentally as the two photon invariant mass distribution would correspond to sharp peaks at the respective $m_X$.

\section{Summary and Conclusions}

We studied the weak radiative decays of $B$ mesons that occur at tree-level in the Standard Model. We presented a numerical estimate for the inclusive double radiative decay $b \to X_c \gamma \gamma$ based on a free quark decay calculation. Our estimate indicates that this mode is about an order of magnitude larger than the double radiative penguin mode $b \to X_s \gamma \gamma$. As such it can be studied at future Super-B factories in experiments designed to observe the double radiative penguin mode, and in fact will be a significant background to that mode.

We reviewed the HQET formalism in connection with the study of the single radiative exclusive modes of  the form $B\to D^\star \gamma$. We first reproduced existing results in the literature for both charged and neutral modes with a leading order calculation including only the $(0^-,1^-)$ doublet as intermediate states. These calculations exhibit a large sensitivity to the value of the electromagnetic couplings due to a partial cancellation. We then improved these lowest order results by including certain known terms of order $(m_b-m_c)/m_c$ as well as by introducing the positive parity doublets $S$ and $T$ as intermediate states. These two ingredients significantly enhance the predictions as they remove the cancellation that occurs at lowest order. The framework is a good approximation in the $m_b\sim m_c$ limit, but significant corrections are expected for the physical values of quark masses. 

Finally we extended the calculations to the double radiative decay modes of the form $B\to D \gamma\gamma$. Once again we found significant cancellations between the lowest order terms and a much larger rate when the positive parity states are included in the calculation. The mode with the most favorable CKM angles is predicted at the $10^{-8}$ level, comparable to predictions for $B\to K \gamma\gamma$. 
 
We expect the HQET formalism to work best in the case when the velocity of the heavy hadrons remains constant during the transition. Whereas this is kinematically impossible in the single radiative decay modes, there are regions of phase space in the double radiative decay modes where this could be tested (in principle at least). We illustrate these regions with plots of the relevant differential decay rates.

\begin{acknowledgments}

This work  was supported in part by DOE under contract 
number DE-FG02-01ER41155. We thank Hai-Yang Cheng and Soeren Prell for discussions.

\end{acknowledgments}

\appendix

\section{Kinematics}

The most general amplitude for decays of the form $M \to V(\eta)\gamma(q,\epsilon)$ consistent with electromagnetic gauge invariance can be written in terms of two form factors. Labeling the momentum of the initial state $M$ with its four velocity in its rest frame, $p=M_Mv$, and denoting the polarization of $V,\gamma$ by $\eta,\epsilon$ respectively
\begin{equation}
{\cal M}(M\to V\gamma)=\epsilon^{\star\mu}\left[ iF_M  \epsilon_{\mu\nu\alpha\beta} v^\nu q^\alpha \eta^{\star\beta}+ F_E\left(  
q\cdot\eta^\star v_\mu-q\cdot v\eta_{\mu}^\star
\right)\right].
\end{equation}
Summing over the photon and vector meson polarizations the partial decay rate is given by 
\begin{equation}
\Gamma(M\to V\gamma)=\frac{E_\gamma^3}{4\pi M_M^2}
\left(|F_M|^2+|F_E|^2\right),
\end{equation}
For the heavy meson formalism in which  the meson fields are  normalized as 
\begin{equation}
<M(v^\prime,k^\prime)|M(v,k)>=2v^0\delta_{vv^\prime}(2\pi)^3\delta^3(\vec{k}-\vec{k}^\prime).
\end{equation}
the decay rate becomes instead
\begin{equation}
\Gamma(M\to V\gamma)=\frac{E_\gamma^3}{4\pi}\frac{E_V}{M_M}
\left(|F_M|^2+|F_E|^2\right),
\end{equation}
where we have used exact kinematics, otherwise in the heavy quark symmetry limit $E_V=M_M$ as well.

The amplitude for decays of the type $M\to M^\prime \gamma \gamma$, with $M, M^\prime$ pseudoscalar mesons, 
\begin{equation}
{\cal M}(M(p) \rightarrow M^\prime (p_3) \gamma(k_1,\epsilon_1)\gamma(k_2,\epsilon_2))
=\epsilon^{\star\mu}(k_1)\epsilon^{\star\nu}(k_2)M_{\mu\nu}(p,k_1,k_2),
\end{equation}
is well known from the kaon literature \cite{Ecker:1987hd}. The most general decay amplitude $M_{\mu\nu}(p,k_1,k_2)$ consistent with electromagnetic gauge invariance and Bose symmetry contains four form factors. We write them here in terms of the velocity of $M$ in its rest frame, so that $p=M_M v$ as it appears within the HQET. 
\begin{eqnarray}
M_{\mu\nu}&=&\frac{A(z,y)}{M_M^2}  \left(k_{2\mu}k_{1\nu}-g_{\mu\nu}k_1 \cdot k_2\right) +
i\frac{C(z,y)}{M_M^2}  \epsilon_{\mu\nu\alpha\beta}k_1^{\alpha}k_2^{\beta}\label{generalff}
\\
          &+& \nonumber  \frac{2B(z,y)}{M_M^2}  \left[ v \cdot k_1 v_{\nu}k_{2\mu}+
v \cdot k_2 v_{\mu}k_{1\nu}-v \cdot k_1 v \cdot k_2 g_{\mu\nu}-
k_1 \cdot k_2 v_{\mu}v_{\nu} \right]\\
&+& \nonumber i \frac{D(z,y)}{M_M^2}  \left[v \cdot k_1 \epsilon_{\mu\nu\alpha\beta}k_2^{\alpha}v^{\beta}+
v \cdot k_2 \epsilon_{\mu\nu\alpha\beta}k_1^{\alpha}v^{\beta}
+ (v_{\mu}\epsilon_{\nu\alpha\beta\gamma}+v_{\nu}\epsilon_{\mu\alpha\beta\gamma})k_1^{\alpha}k_2^{\beta}v^{\gamma}\right],
\end{eqnarray}
where
\begin{equation}
y=\frac{v \cdot (k_1-k_2)}{M_M}, \hspace{5mm}  z=\frac{(k_1+k_2)^2}{M_M^2},\hspace{5mm} r=\frac{M_{M^\prime}}{M_M}.
\label{kindef}
\end{equation}
The relation between these dimensionless variables and the energy of the two photons is
\begin{eqnarray}
E_1&=&\frac{1}{4} M_M ( (z+2y+1)- r^2 )\nonumber \\
E_2&=&\frac{1}{4} M_M ( (z-2y+1)- r^2 ).
\end{eqnarray}
Recently, Hiller and Safir \cite{Hiller:2004wc} have claimed in the context of 
$B \to K \gamma \gamma$ that there are three additional form factors,
\begin{eqnarray}
M_{\mu\nu}^\prime&=&
  i \frac{C^+(z,y)}{M_M^3}  \left[k_1 \cdot k_2 \epsilon_{\mu\nu\alpha\beta}
(k_1^{\alpha}+k_2^{\beta})v^{\beta}
+ (k_{2\mu}\epsilon_{\nu\alpha\beta\gamma}+k_{1\nu}\epsilon_{\mu\alpha\beta\gamma})k_1^{\alpha}k_2^{\beta}v^{\gamma}\right]\label{hiller}
\\
          &+& \nonumber i \frac{C^-(z,y)}{M_M^3}  \left[k_1 \cdot k_2 \epsilon_{\mu\nu\alpha\beta}
(k_1^{\alpha}-k_2^{\alpha})v^{\beta}
- (k_{2\mu}\epsilon_{\nu\alpha\beta\gamma}-k_{1\nu}\epsilon_{\mu\alpha\beta\gamma})k_1^{\alpha}k_2^{\beta}v^{\gamma}\right]\\
          &+& \nonumber i \frac{D^-(z,y)}{M_M^2}  \left[v \cdot k_1 \epsilon_{\mu\nu\alpha\beta}k_2^{\alpha}v^{\beta}-
v \cdot k_2 \epsilon_{\mu\nu\alpha\beta}k_1^{\alpha}v^{\beta}
+ (v_{\mu}\epsilon_{\nu\alpha\beta\gamma}-v_{\nu}\epsilon_{\mu\alpha\beta\gamma})k_1^{\alpha}k_2^{\beta}v^{\gamma}\right]. 
\end{eqnarray}
It is well known, however, that these form factors are not independent and can be reduced to the ones in Eq.~(\ref{generalff}). This follows from the fact that in four dimensions there are at most four linearly independent four vectors, and this gives rise to the Schouten identity \cite{Fearing:1994ga}. In this case all three terms in Eq.~(\ref{hiller}) reduce to the second form factor in Eq.~(\ref{generalff}) so that \footnote{The authors of Ref.\cite{Hiller:2004wc} agree with us and have added a note to this effect in their paper.}
\begin{eqnarray}
M_{\mu\nu}^\prime =
 \left(M_M y C^+(z,y) +    \frac{M_M}{2}\left(1+z-r^2\right)C^-(z,y) - D^-(z,y)\right) \epsilon_{\mu\nu\alpha\beta}k_1^{\alpha}k_2^{\beta}
\end{eqnarray}

The physical region in the dimensionless variables z and y is 
given by
\begin{equation}
0 \leq |y| \leq\frac{1}{2} \lambda^{1/2}(1,r^2,z), \hspace{5mm} 0 \leq z \leq (1-r)^2,
\end{equation}
with
\begin{equation}
\lambda(a,b,c)=a^2+b^2+c^2-2(ab+ac+bc). 
\end{equation}

Note that the invariant amplitudes $A(z,y)$, $B(z,y)$ and $C(z,y)$ have to be symmetric under the
interchange of $k_1$ and $k_2$ as required by Bose symmetry, while $D(z,y)$ is antisymmetric.
Using the definitions (\ref{generalff}) the double differential rate for unpolarized photons and conventionally normalized meson fields is given by (in the rest frame of $M$)
\begin{eqnarray}
\frac{\partial^2{\Gamma}}{\partial{y} \partial{z}}&=&\frac{M_M}{2^9\pi^3}
\hspace{2mm}[z^2( \hspace{2mm}|A+B|^2+|C|^2 \hspace{2mm} ) \label{diffdis} \\
     &+& \nonumber \hspace{3mm} (\hspace{1mm}|B|^2+|D|^2 \hspace{1mm})\hspace{2mm}
(y^2-\frac{1}{4}\lambda(1,r^2,z))^2].
\end{eqnarray}

With HQET normalization for the meson fields and with the two photons retaining their usual normalization Eq.~(\ref{diffdis}) is replaced by
\begin{eqnarray}
\frac{\partial^2{\Gamma}}{\partial{y} \partial{z}}&=&\frac{M^2_ME_{M^\prime}}{2^9\pi^3}
\hspace{2mm}[z^2( \hspace{2mm}|A+B|^2+|C|^2 \hspace{2mm} )  \\
     &+& \nonumber \hspace{3mm} (\hspace{1mm}|B|^2+|D|^2 \hspace{1mm})\hspace{2mm}
(y^2-\frac{1}{4}\lambda(1,r^2,z))^2],
\label{diffdishq}
\end{eqnarray}
of course, in the heavy quark limit $E_{M^\prime}\to M^\prime$ as well.

\end{document}